\documentclass{astlb}

\usepackage[T2A]{fontenc}
\usepackage[utf8]{inputenc}
\usepackage[russian,english]{babel}

\usepackage{lscape}
\usepackage{booktabs} 
\usepackage{multicol,lipsum}
\usepackage{xspace} 

\usepackage{ragged2e}
\usepackage{color}

\usepackage{graphicx}	
\usepackage{amssymb}
\usepackage{amsmath}	
\usepackage{longtable}

\def\ps1{{Pan-STARRS1}}

\def\20{{L20}}

\def\nu{NuSTAR}

\def\telesc{IBIS}

\def\ths{{\hspace{2pt}}}
\def\dhs{{\hspace{3pt}}}

\begin{document}

\journalinfo{(2022)}{48}{11}{680}[687]

\title{Search for Nonthermal X-ray Emission in the Ophiuchus Galaxy Cluster}

\author{
  Roman A. Krivonos\address{1}\email{krivonos@cosmos.ru}
\addresstext{1}{\it Space Research Institute, Russian Academy of Sciences, Moscow, 117997 Russia}
}

\shortauthor{Krivonos et al.}

\shorttitle{Search for Nonthermal X-ray Emission in the Ophiuchus Galaxy Cluster}

\submitted{July 1, 2022; revised December 2, 2022; accepted December 2, 2022}

\begin{abstract}
We present the results of our study of the X-ray emission from the Ophiuchus galaxy cluster based on INTEGRAL/IBIS data in the energy range $20-120$~keV. Our goal is the search for a nonthermal emission component from the cluster. Using the INTEGRAL data over the period of observations 2003-2009, we have constructed the images of the Ophiuchus galaxy cluster in different energy bands from 20 to 120~keV with the extraction of spectral information. We show that in the hard X-ray energy band the source is an extended one with an angular size of $4.9'\pm0.1'$. Assuming a fixed intracluster gas temperature of 8.5~keV, a power-law component of the possible nonthermal X-ray emission is observed at a $5.5\sigma$ significance level, the flux from which is consistent with previous studies. However, in view of the uncertainty in constraining the thermal emission component in the X-ray spectrum at energies above 20 keV, we cannot assert that the nonthermal emission of the cluster has been significantly detected. Based on the fact of a confident detection of the cluster up to 70~keV, we can draw the conclusion only about the possible presence of a nonthermal excess at energies above 60~keV.


\end{abstract}

	\section{INTRODUCTION}
	
Most of the galaxies are in gravitationally bound systems-galaxy groups and clusters that can include from a few to a few thousand galaxies. In galaxy clusters the galaxies themselves account for only 1\% of the total mass, which is determined mainly by dark matter (90\%) and a hot gas (9\%) with a temperature of $10^7-10^8$~K. Owing to plasma bremsstrahlung, galaxy clusters are bright X-ray sources.

Massive clusters are usually formed through the merger of smaller ones and, as a consequence, intracluster-gas-heating shocks emerge. According to theoretical models, relativistic electrons emitting in the X-ray \citep{2003MNRAS.342.1009M} and visible \citep{2015MNRAS.453.1990Y} bands through inverse Compton scattering and in the radio band through the synchrotron mechanism \citep{2020PhRvL.124e1101B} are produced during such mergers as a result of first- and second-order Fermi acceleration. There are also other models that explain the origin of the nonthermal emission in the radio and X-ray bands by the annihilation of darkmatter particles \citep{2016JCAP...11..033M}.

The Ophiuchus galaxy cluster is one of the brightest in the X-ray energy band. The X-ray emission from it was first detected in the early 1980s by \cite{1980BAAS...12..487J,1981ApJ...245..799J} based on data from the A-2 instrument onboard the HEAO-1 observatory. This allowed one to measure the hot gas temperature, $kT=8\pm2$, and to estimate the angular size of the cluster, $R_{\mbox{\scriptsize{Oph}}}$ $=$ 3.9 $^{+{\ths}\textrm{1.2}}_{-{\ths}\textrm{1.4}}$~arcmin (90\% significance level). Its redshift $z = 0.028 \pm 0.003$ was also measured from the shift of absorption lines. \cite{1981PASJ...33...57W} independently detected the Ophiuchus cluster in the optical wavelength range using images on photographic plates taken with the Palomar Schmidt telescope. They plotted the dependence of the density of galaxies in the cluster $\rho(r)$ on distance $r$ from the cluster center and found that $\rho(r) \sim (1 + r^{{\ths}2}/r_c^{{\ths}2})^{-1}$, where the core radius $r_c$ $=$ 9.5$'$ $\pm$ 1.5$'$. They also showed that the Ophiuchus cluster recedes from the Earth with a velocity $v\sim7400$ km~s$^{-1}$ and concluded that the cluster is approximately at the same distance from us as the Coma cluster. Hence, the redshift is found to be $z\sim$~0.025, consistent with the result of \cite{1981ApJ...245..799J}.
	
Based on EXOSAT data, \cite{1987MNRAS.227..241A} estimated the plasma temperature in the cluster to be $kT=$ 9.4$^{+{\ths}\textrm{1.5}}_{-{\ths}\textrm{1.2}}$ and the ion abundance to be 0.26$^{+{\ths}\textrm{0.14}}_{-{\ths}\textrm{0.13}}$ in solar units (90\% significance level). \cite{1996PASJ...48..565M} fitted the X-ray emission from the cluster by a thermal bremsstrahlung model with two iron lines, K$_\alpha$ and K$_\beta$. Based on data from the gas imaging spectrometer (GIS) onboard the ASCA satellite, they obtained the following emission parameters: z $=$ 0.031$^{+{\ths}\textrm{0.004}}_{-{\ths}\textrm{0.007}}$, kT $=$ 9.7 $\pm$ 0.4~keV, the iron K$_\alpha$ and K$_\beta$ $-$ 6.60 $\pm$ 0.04~keV and 7.8 $\pm$ 0.2~keV, respectively, and the iron abundance 0.24 in solar units. The $\beta$-model was chosen to fit the gas density distribution in the cluster:

	\begin{equation}
	\label{bet_mod}
	n(r) = n_0\left[1 + \left(\frac{r}{r_c}\right)^2\right]^{-3{\beta}/2}
	\end{equation}

\noindent where $n_{0}$ is the gas density at the cluster center and r is the distance from the cluster center. The following model parameters were obtained: $r_c$ $=$ 3.8$'$, $\beta$ $=$ 0.62. 

A study of the possible contribution of nonthermal
emission to the X-ray spectrum of the Ophiuchus
cluster (and other clusters) is greatly complicated by
the uncertainty in measuring the thermal emission
component of the hot gas, which has a multitemperature
structure and contains a central cooling flow.
In view of the relatively high cluster temperature,
measurements at energies above 10~keV are critically
important for constraining the parameters of the cluster
gas thermal emission model.

\cite{Nevalainen2004} announced the detection of nonthermal emission from the Ophiuchus cluster based on BeppoSAX data at a $2\sigma$ significance level. The plasma temperature in the cluster derived by them is kT $=$  9.1$^{+{\ths}\textrm{0.6}}_{-{\ths}\textrm{0.5}}$~keV. The INTEGRAL gamma-ray observatory gives a unique opportunity to obtain a broadband X-ray spectrum of the Ophiuchus cluster. Based on data from the JEM-X and IBIS/ISGRI instruments onboard this observatory, \cite{Eckert2008} recorded nonthermal emission from the cluster with a $4-6\sigma$ significance. They showed that the cluster is detected with confidence at energies above 20~keV. In addition to this study, \cite{Nevalainen2009} extended the spectral coverage to energies below 10~keV using XMM-Newton data and performed a detailed spatial analysis of the thermal emission from the cluster.

The goal of this paper is to continue the search for nonthermal emission from the Ophiuchus cluster and to constrain its parameters using INTEGRAL data in a wide energy range, from 20 to 120 keV. During the long-term in-orbit operation the characteristics of the INTEGRAL instruments gradually degraded, especially after 2009. More specifically, from 2009 to 2012 the energy resolution of the ISGRI detector deteriorated from 5 to 15 keV, while the lower boundary of the energy range rose from 17 to 23 keV \citep{2013arXiv1304.1349C}. The degradation of the IBIS detector state is evidenced by the observed drop in the count rate from a ``standard candle'' in X-ray astronomy, the Crab Nebula, by ${\sim}20\%$ in the period from 2009 to 2010 \citep[see Fig.~6 in][]{2021A&A...651A..97N}. This greatly complicates the construction of an average Xray spectrum over a long time interval. Therefore, in this paper we use the highest-quality IBIS data obtained before 2009. Compared to \cite{Eckert2008}, we use not only a larger volume of data (the observational data before 2007 were included in Eckert’s study), but also more up-to-date energy calibrations and systematic noise suppression methods in IBIS images \citep{Krivonos2010}.

	\begin{figure}
		\begin{minipage}{1\linewidth}
			\center{\includegraphics[width=0.99\linewidth]{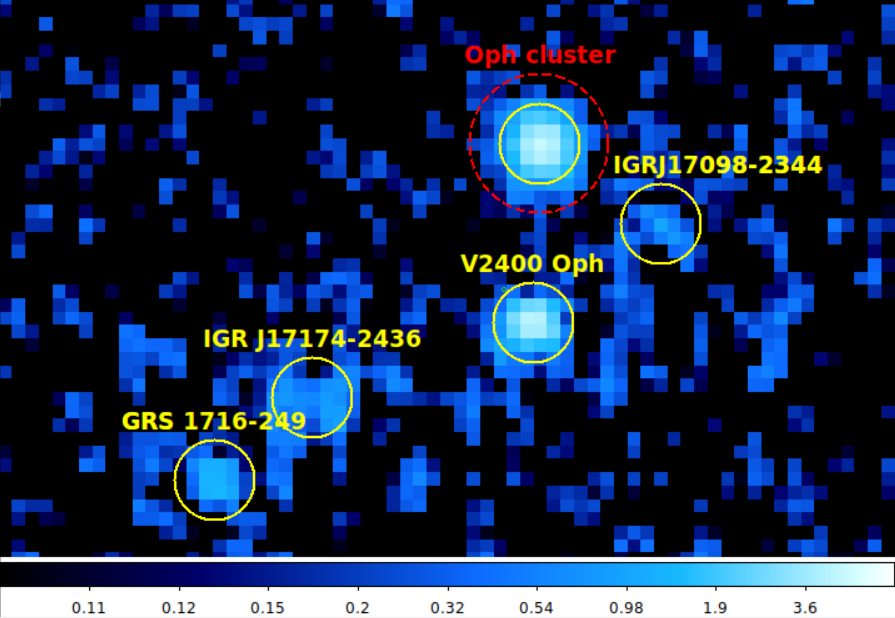}}
			\caption{\small{IBIS image of the Ophiuchus galaxy cluster region in the $20-60$~keV energy band (in mCrab). The yellow circles mark the positions of the known X-ray sources in the field of view, the circle size corresponds to the angular resolution of the telescope (12$'$). The red dashed circumference indicates the 20$'$ region that was used to construct the radial cluster profile. The pixel size is 4$'$.}}
			\label{fig:oph}
		\end{minipage}
	\end{figure}

\section{OBSERVATIONS AND DATA ANALYSIS}
\label{sec:data}
The INTEGRAL (INTErnational Gamma-Ray Astrophysics Laboratory) gamma-ray observatory is designed to study cosmic objects in the hard X-ray and gamma-ray bands \citep{2003A&A...411L...1W}. The observatory is a joint project of the European Space Agency (ESA), Roskosmos, and NASA. It was placed into a highly elliptical geosynchronous orbit in 2002 with a Proton launcher.

In this paper we use data from the IBIS telescope with the working energy range 15 keV -- 10 MeV and an angular resolution of 12$'$ in a 28$^{\circ}$ $\times$ 28$^{\circ}$  field of view. The telescope contains two layers of detectors: the upper one, ISGRI, consisting of 4 $\times$ 4 $\times$ 2~mm CdTe pixels sensitive at energies up to 1 MeV \citep{2003A&A...411L.141L} and the lower one, PICsIT, consisting of 9 $\times$ 9 $\times$ 30~mm CsI pixels operating at energies above 150~keV.

The IBIS telescope operates on the principle of a coded aperture. Passing through a mask located at some height from the detector, the emission from different sources creates a superposition of shadows (shadowgram) on the detector. The sky image is reconstructed by cross-correlating the shadowgram with the mask transmission function; as a result, an image of the observation region on the celestial sphere is obtained.
	
The IBIS data were analyzed with the software developed at the Space Research Institute of the Russian Academy of Sciences \citep[see][]{Krivonos2010,Krivonos2012,Krivonos2017,2014Natur.512..406C}. We constructed the images of the cluster region in several bands from 20 to 120 keV based on the data from 2003 to 2009. The data set being used includes 7367 individual observations of the central part of the Galaxy with an angular shift no more than 15 deg. from the cluster position. The total exposure time is 18 Ms (or, given the dead time, 13Ms). We use the energy calibrations of the INTEGRAL Offline Scientific Analysis (OSA) software of version 10.2.

\subsection{Angular Size of the Cluster}
\label{rad_prof_sec}

The X-ray emission from the Ophiuchus galaxy cluster is known to be an extended one with a characteristic size of about 15 arcmin \citep{Nevalainen2009,Werner2016}, which is comparable to the angular resolution of the IBIS telescope (12 arcmin). Despite the fact that the coded-aperture telescopes are unable to construct the images of extended objects, some information about their angular sizes can still be obtained from the IBIS data \citep[see, e.g.,][]{2019MNRAS.489.1828K}. In addition, special methods different from those that are used for point sources are required to be applied to properly extract the flux from an extended object. In particular, for the flux from an X-ray source to be correctly estimated, the convolution function, when reconstructing its image, must maximally coincide with the profile of this source for a given telescope.
	
\begin{figure}
		\begin{minipage}{1\linewidth}
			\center{\includegraphics[width=0.99\linewidth]{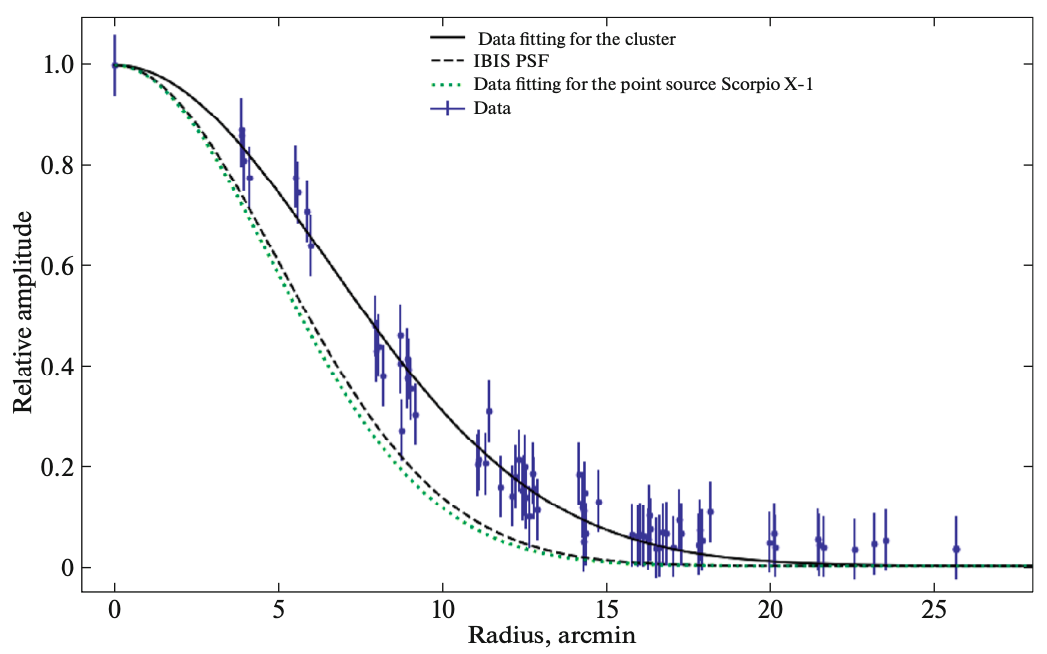}}
			\caption{\small{
			The radial brightness profile of the Ophiuchus cluster in the $20-60$~keV energy band from ISGRI data (blue dots). The black dashed line indicates the IBIS PSF, the green dotted line indicates the fit to the radial profile of the point source Scorpio~X-1. The solid black line indicates the resulting profile from the IBIS PSF and the cluster emission profile. To avoid the coalescence of dots in the figure, we added an artificial scatter of their horizontal coordinates at a $\pm0.3'$ level.}}
			\label{fig:prof}
		\end{minipage}
	\end{figure}

To determine whether the Ophiuchus cluster is a point or extended source for the IBIS telescope, we investigate the radial cluster flux profile. Figure~\ref{fig:oph} shows a map of the flux from the Ophiuchus cluster and the X-ray sources nearest to it in the 20–60 keV energy band. To construct the radial profile, we took into account the pixels lying within a circle of radius 20$'$. At such a distance from the cluster center the intensity drops almost to zero, as can be seen from the radial profile in Fig.~\ref{fig:prof}.

Although the cluster X-ray brightness profile is described by the $\beta$-model, to check whether the emission is an extended one, it will suffice to fit the distance dependence of the flux by a Gaussian:

\begin{equation}\label{intens}I(r) \sim \exp\left(- \ln(2)\frac{r^2}{R^2}\right),
\end{equation} 

where $r$ is the distance from the cluster center and $R$ is the half width at half maximum (HWHM) for the function $I(r)$. In this way we find that $R=7.7'\pm0.1'$ \mbox{($\chi^2$/d.o.f = 37.7 / 75 = 0.50)}. The value of $R$ corresponds to a superposition of the point spread function (PSF) of the IBIS telescope and the radial profile of the source: $R^2 = R_{\mbox{\scriptsize{\telesc}}}^2 + R_{\mbox{\scriptsize{Oph}}}^2.$. Hence, knowing the HWHM of the IBIS PSF ($R_{\mbox{\scriptsize{\telesc}}}$ $=$ 5.9$'$), we can estimate the HWHM of the source profile: $R_{\mbox{\scriptsize{Oph}}} = 4.9' \pm 0.1'$. 

Performing similar operations for the point X-ray source Scorpio X-1, we obtain $R_{Sco{\dhs}X-1}$ $=$ 5.7$'$ $\pm$ 0.1$'$, corresponding to the HWHM of the IBIS PSF: $R_{\mbox{\scriptsize{\telesc}}}$ $=$ 5.9$'$ (see Fig.~\ref{fig:prof}). Therefore, this source may be deemed a point one.

Comparing the HWHMs of the radial brightness profiles for the Ophiuchus cluster and the point source, we conclude that the cluster is an extended hard X-ray source. In what follows, the X-ray flux from the cluster will be extracted from the maps convolved with a Gaussian function with $R=7.7'$, unless stated otherwise.

\subsection{Model of the Emission from the Cluster}
\label{modRadiation}

The emission model is a sum of the thermal and nonthermal emissions. The APEC model \citep[see, e.g.,][]{2014A&A...562A..11I} is used for the thermal emission. This model fits the spectrum of a collisionally ionized plasma. It is defined by four parameters: the plasma temperature $kT$ (keV), the redshift $z$, the normalization $norm_{\mbox{\small{apec}}}$ (cm$^{-5}$), and the metal abundance in solar units. The nonthermal emission component is fitted by a power-law model. It includes the following parameters: the exponent $\alpha$ and the normalization $norm_{\mbox{\small{pow}}}$ at an energy of 1~keV (photons keV$^{-1}$~cm$^{-2}$~s$^{-1}$).

Thus, there are a total of six unknown parameters. Four of them are fixed from \cite{Eckert2008}: the slope of the power-law model $\alpha=2$, the abundance of 0.49 in solar units, the temperature $kT=8.5$~keV, and the redshift $z = 0.028$. These values are consistent with other studies\footnote{Nevertheless, it should be noted that the assumption about a fixed cluster temperature is a strong one. A mixture of gas phases with temperatures from 1 to 11 keV is known to be present in the central region of the cluster \citep{Werner2016}.} \citep[see, e.g.,][]{1996PASJ...48..565M,Nevalainen2004}. We will fit the two remaining unknown parameters, the normalizations of both emission components normpow and normapec, from the condition for the minimum of the $\chi^2$ statistic.

\subsection{Estimation of the Emission Model Parameters}
\label{mo_pars}
	
As the fluxes from the cluster and the corresponding errors for each of the energy bands we take the values of the convolved flux maps and their errors at the cluster center. 

\begin{figure}[ht]
\begin{minipage}{1\linewidth}
\center{\includegraphics[width=0.9\linewidth]{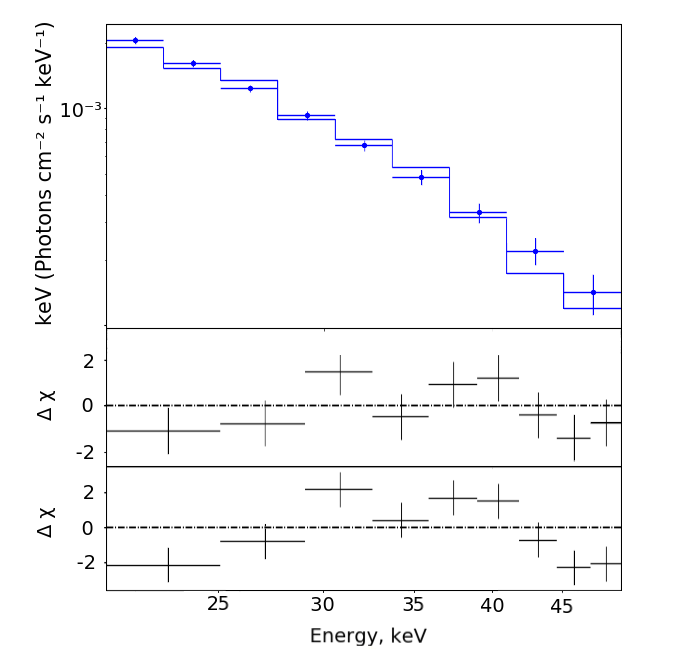}}			
\caption{\small{\textit{Top}: the energy spectrum of the emission from the Ophiuchus cluster from IBIS data in the $20-50$~keV energy band with errors (blue bins) and the APEC thermal spectral model (blue line). \textit{Bottom}: the residuals \{$\Delta \chi$ = (data $-$ model) / error\} of the thermal (APEC; top) and power-law (bottom) spectral models.}}
			\label{th50}	
		\end{minipage}
	\end{figure}

As a result, the statistic takes a minimum value equal to $\chi^2$/d.o.f.=1.9 (d.o.f. corresponds to the number of degrees of freedom) if we set the following values of the parameters being estimated: norm$_{\mbox{\small{pow}}} =$ (3.9 $\pm$ 0.7) $\times$ 10$^{-3}$ photons keV$^{-1}$ cm$^{-2}$ s$^{-1}$ and norm$_{\mbox{\small{apec}}} = 0.28\pm0.01$ cm$^{-5}$. To construct the confidence intervals of both normalizations, we fixed one of these parameters and varied the second one so that the statistic changed by $\delta\chi^2$ = 1. The derived values agree well with previous studies. Our model gives a nonthermal flux of $(6.6\pm1.2)\times10^{-12}$~erg~cm$^{-2}$~s$^{-1}$ in the $20-60$ keV energy band. \cite{Eckert2008} obtained a flux in the $20-60$~keV energy band of $(8.2\pm1.3)\times10^{-12}$~erg~cm$^{-2}$~s$^{-1}$, in agreement with our value within the error limits.

\section{DISCUSSION}
\label{sec:discussion}

To check how well the spectrum at energies below 50~keV is described by the thermal law, it was initially fitted by a power law (Fig.~\ref{th50}). A curvature of the residuals is observed, suggesting that the spectrum is thermal in nature. Indeed, if we now model the spectrum by the APEC thermal model (Fig.~\ref{th50}), in which the temperature and the normalization are free parameters, while the remaining two parameters are the same as those in Subsection 2.3, then the reduced chi-squared will decrease compared to the previous case from 3.65 to 1.28.

\begin{figure*}[ht]
\begin{minipage}{1\linewidth}
\center{\includegraphics[width=0.9\linewidth]{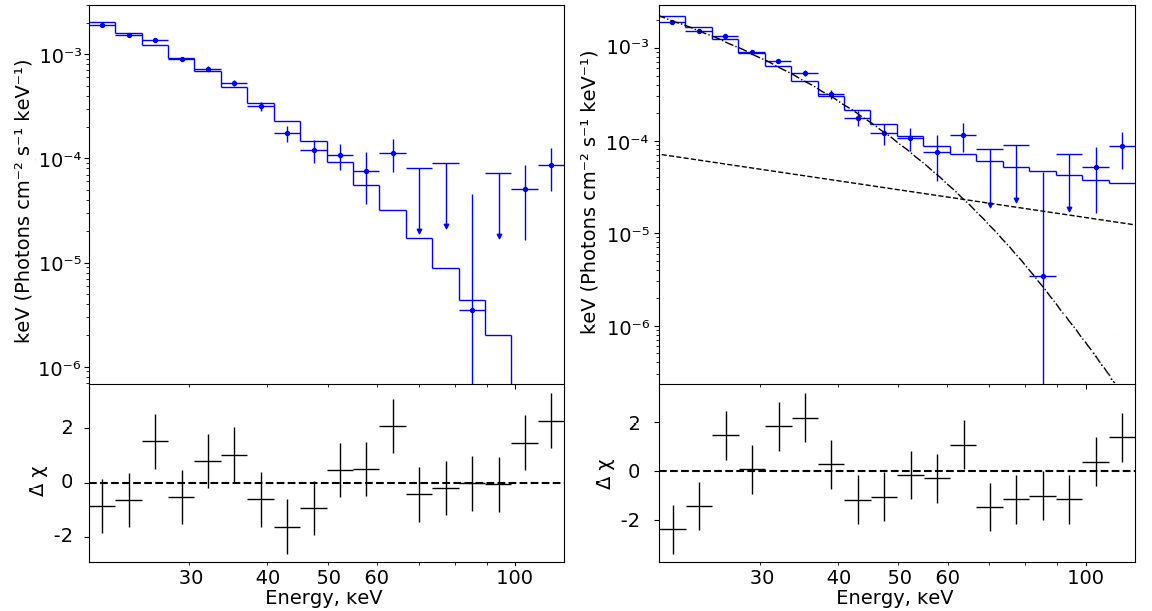}}
\caption{\small{The energy spectrum of the emission from the Ophiuchus cluster from IBIS data in the $20-120$~keV energy band. \textit{Top left}: the spectrum was fitted by the APEC thermal model with a temperature of 11.3~keV (blue histogram). \textit{Top right}: the spectrum was fitted by the sum (blue histogram) of the APEC thermal model with a temperature of 10.8~keV (dash–dotted line) and the power-law nonthermal model (dashed line). \textit{Bottom}: the residuals of both models.}}
		\label{allBins}	
	\end{minipage}
	\end{figure*}

If we now add the remaining bins with energies above 50~keV to our analysis (Fig.~\ref{allBins}, left panel), then the residuals will hardly change. The temperature is determined at a level of ${\approx}11.3$~keV. Then, we will add a nonthermal component obeying a power law to the model (Fig.~\ref{allBins}, right panel). The statistic becomes equal to 1.37. In attempting to construct the confidence interval for the normalization of the nonthermal component norm$_{\mbox{\small{pow}}}$, it is possible to obtain
only an upper limit on it, which is $2.3 \times 10^{-3}$ photons~keV$^{-1}$~cm$^{-2}$~s$^{-1}$.

Then, we will set the temperature $T = 8.5$~keV, as was done in \cite{Eckert2008}. The 1$\sigma$ confidence interval for norm$_{\mbox{\small{pow}}}$ in this case is $(3.2 - 4.6) \times 10^{-3}$, giving a 5.5$\sigma$ detection significance for the nonthermal signal. The statistic becomes equal to $\chi^2$/d.o.f. = 1.9. The contributions from the thermal and non- thermal emission components in the $20-60$~keV energy band at this temperature are related as 4:1.

Their total flux in this energy band is $(3.7 \pm 0.25) \times 10^{-11}$~erg~cm$^{-2}$~s$^{-1}$. The nonthermal flux in this band (6.6 $\pm$ 1.2) $\times 10^{-12}$~erg~cm$^{-2}$~s$^{-1}$. The nonthermal flux at energies $50-70$~keV accounts for 5.4\% of the total flux in the $20-70$~keV energy band.
	
\begin{figure}[ht]
\begin{minipage}{1\linewidth}
\center{\includegraphics[width=0.99\linewidth]{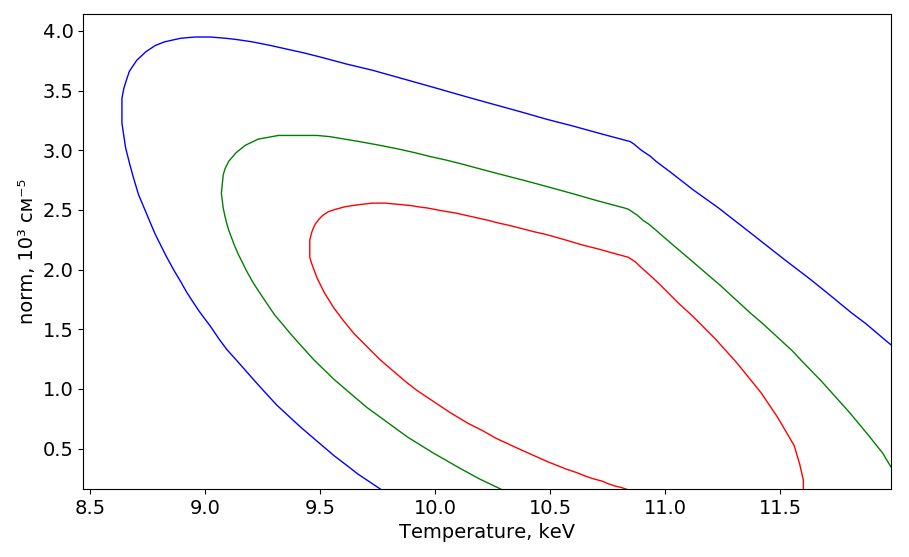}}
\caption{\small{Two-dimensional $1\sigma$, $2\sigma$ and $3\sigma$ confidence contours (with two free parameters) for the normalization of the nonthermal component and the temperature of the thermal component in the $20-120$~keV energy band.}}
\label{contour_nT}	
\end{minipage}
\end{figure}

A correlation between the temperature of the thermal component and the normalization of the nonthermal component is also observed: as the temperature rises from 8.5~keV to its optimal value of $T\approx10.8$~keV, norm$_{\mbox{\small{pow}}}$ drops from ($3.9 \pm 0.7) \times 10^{-3}$ cm$^{-5}$ to $1.3^{\hspace{2pt}+0.9}_{\hspace{2pt}-1.3}$ $\times 10^{-3}$ cm$^{-3}$. The correlation is presented in more detail in Fig.~\ref{contour_nT}. Thus, because of the significant uncertainty in the parameters of the thermal emission component in the X-ray spectrum, it is impossible to talk with confidence about a significant detection of nonthermal emission from the cluster. We can make the statement only about the possible presence of a nonthermal excess at energies above 60~keV.

\begin{figure*}[ht]
	\begin{minipage}{1\linewidth}
		\center{\includegraphics[width=1\linewidth]{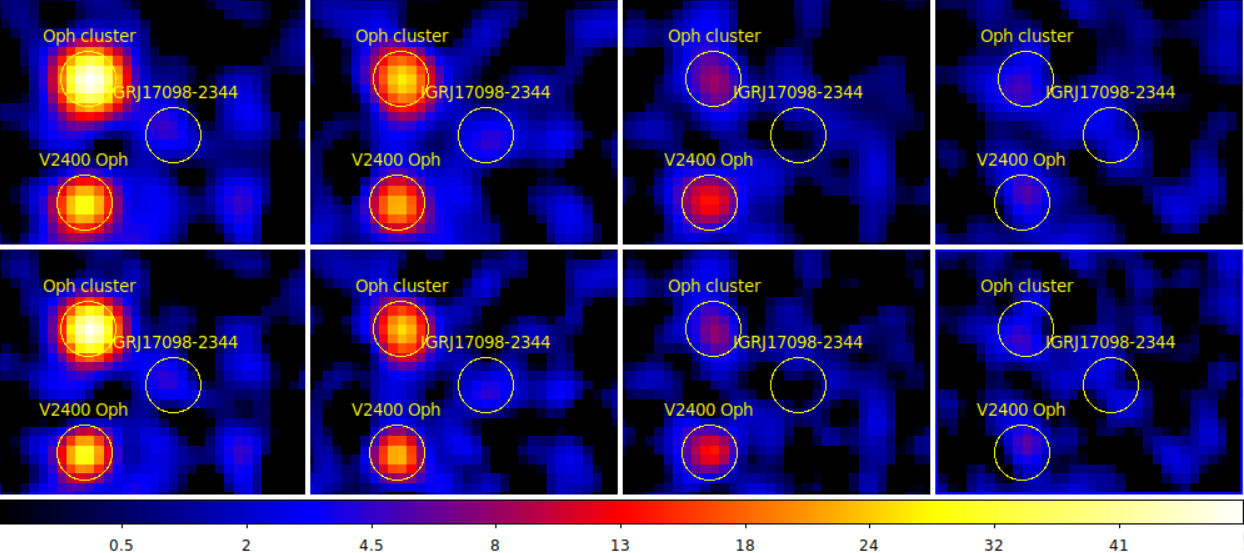}}
		\caption{\small{Significance maps of the signal from the Ophiuchus cluster and the sources nearest to it. Four energy bands from left to right are presented in each of the two rows: $20-30$, $30-40$, $40-50$ and $50-70$~keV, respectively. \textit{Top}: the maps were obtained by convolving the flux maps and their errors with a Gaussian with R = 7.7$'$ (see Subsection~\ref{rad_prof_sec}). \textit{Bottom}: Similar to the upper row of maps, but for a Gaussian with R = 5.9$'$, corresponding to a point source.}}
		\label{maps_sign}	
	\end{minipage}
\end{figure*}

The cluster significance maps in four bands from 20 to 70 keV are presented in Fig.~\ref{maps_sign}. In the $50-70$~keV energy band the cluster is observed with a $4.3\sigma$ significance, i.e., we can talk about a significant detection of the cluster at energies up to 70~keV.

\section{CONCLUSIONS}
\label{sebsec:4}
In this paper we presented the results of the observations of the Ophiuchus galaxy cluster with the IBIS telescope onboard the INTEGRAL observatory in a wide energy band, $20-120$~keV, based on the observational data from 2003 to 2009. The source was shown to be an extended one in the hard X-ray band.

We demonstrated the presence of an excess nonthermal emission with respect to the thermal one with a fixed temperature of 8.5 keV at a 5.5$\sigma$ significance level. In that case, the normalization of the power-law nonthermal emission model is equal to $(6.6\pm1.2)\times10^{-12}$~erg~cm$^{-2}$~s$^{-1}$ in the $20-60$~keV energy band, consistent with the result of \cite{Eckert2008} within the error limits (see Subsection~\ref{modRadiation}). However, because of the uncertainty in the parameters of the thermal emission component in the X-ray spectrum at energies above 20~keV, it is impossible to make the statement about a significant detection of nonthermal emission from the cluster. Based on the fact of a confident detection of the cluster up to 70~keV, we can draw only the conclusion about the possible presence of a nonthermal excess at energies above 60~keV. For a confident answer to this question, it is necessary to improve the quality of observations at energies above 60~keV, which can be done using the entire INTEGRAL observational data set. We are planning to do this in our future paper.

On the other hand, it is required to supplement the spectral coverage of the hard energy band by measurements below 10~keV, data from telescopes with a sufficient spatial resolution for a more severe limitation of the multitemperature thermal model of the cluster, for example, as was done in \cite{Nevalainen2009}. This will allow one to avoid using the assumption about a fixed cluster temperature. According to \cite{Werner2016}, a hot gas with a temperature up to 11~keV is observed in the Ophiuchus cluster. The emission of a gas with such a temperature will dominate at energies above~20 keV, which is observed in this paper: fitting the X-ray spectrum with a free temperature gives an estimate of about 11~keV (see Section~\ref{sec:discussion}).

However, adding the X-ray data below 10~keV to the IBIS spectrum involves difficulties. Most of the X-ray telescopes have small fields of view, which does not allow a homogeneous study of the entire cluster to be carried out. As a result, the multitemperature thermal emission will be measured only in a specific part of the cluster, whereas the IBIS coded-aperture telescope onboard the INTEGRAL observatory measures the emission from the cluster though entirely, but with an uncertain region of the signal extraction that is determined only by the angular resolution of the telescope (12$'$). A more or less unambiguous correspondence to the measured mixture of spatial different-temperature components in the cluster by the INTEGRAL observatory may be expected when using data from the Mikhail Pavlinsky ART-XC telescope onboard the Spektr-RG observatory, which provides a complete coverage of the cluster and has a sufficiently wide operating energy range, $4-30$~keV.

\section{ACKNOWLEDGMENTS}

I am grateful to S.Yu.~Sazonov, A.A.~Vikhlinin, and E.M.~Churazov for the recommendations to improve the paper.  This work is based on observations with INTEGRAL, an ESA project with instruments and the science data centre funded by ESA member states (especially the PI countries: Denmark, France, Germany, Italy, Switzerland, Spain), and Poland, and with the participation of Russia and the USA.

\section{FUNDING}

This work was supported by the Russian Science Foundation within scientific project no. 19-12-00396.

\section{CONFLICT OF INTEREST}

The authors of this work declare that they have no conflicts of interest.



\bibliographystyle{astl}
\bibliography{biblio} 


{\it Translated by V. Astakhov}

{\it Latex style was created by R. Burenin}

\label{lastpage}

\end{document}